# Photonics-assisted wideband RF self-interference cancellation with digital domain amplitude and delay pre-matching

Taixia Shi[a,b], Moxuan Han[a,b] and Yang Chen[a,b,*]

[a] Shanghai Key Laboratory of Multidimensional Information Processing, East China Normal University, Shanghai, 200241, China
[b] Engineering Center of SHMEC for Space Information and GNSS, East China Normal University, Shanghai, 200241, China
[*] ychen@ce.ecnu.edu.cn

**ABSTRACT**
A photonics-based digital and analog self-interference cancellation approach for in-band full-duplex communication systems and frequency-modulated continuous-wave radar systems is reported. One dual-drive Mach-Zehnder modulator is used to implement the analog self-interference cancellation by pre-adjusting the delay and amplitude of the reference signal applied to the dual-drive Mach-Zehnder modulator in the digital domain. The amplitude is determined via the received signal power, while the delay is searched by the cross-correlation and bisection methods. Furthermore, recursive least squared or normalized least mean square algorithms are used to suppress the residual self-interference in the digital domain. Quadrature phase-shift keying modulated signals and linearly frequency-modulated signals are used to experimentally verify the proposed method. The analog cancellation depth is around 20 dB, and the total cancellation depth is more than 36 dB for the 2-Gbaud quadrature phase-shift keying modulated signals. For the linearly frequency-modulated signals, the analog and total cancellation depths are around 19 dB and 34 dB, respectively.

**Keywords:** Microwave photonics; self-interference cancellation; in-band full-duplex; FMCW radar.

## 1. Introduction

With the development of modern radio technology, the bandwidth and data transmission rate of radio applications are getting larger and larger. However, spectrum resources are limited, so there is a great need to develop new spectrum resources and use the existing spectrum resources efficiently. Compared with the existing time division duplex and frequency division duplex wireless communication systems, the in-band full-duplex (IBFD) communication systems can simultaneously transmit and receive signals at the same frequency [1-2], which can effectively improve the efficiency of time or spectrum utilization. However, the IBFD communication system will cause a strong self-interference (SI) because of simultaneous transmitting and receiving signals at the same frequency, and it is difficult to remove the SI directly through a notch filter or a bandpass filter. Besides, compared with the conventional pulse radar, the frequency-modulated continuous-wave (FMCW) radar can work at a lower cost and lower power consumption with a larger time-bandwidth product [3-4], which improves the efficiency of time and energy utilization. However, the FMCW radar also suffers from the SI from the transmitting antenna to the receiving antenna. Therefore, self-interference cancellation (SIC) is particularly important for the IBFD communication systems and the FMCW radar systems.

Since the SI is much stronger than the signal of interest (SOI), the elimination of SI is often realized through several steps [5], mainly including the antenna domain cancellation, analog domain cancellation, and digital domain cancellation. The analog cancellation cancels the SI by constructing a reference signal that is similar to the SI. There are many electrical-based SIC methods [5], but these methods can handle limited bandwidth, which is still hard to meet the needs of radio applications with large bandwidth and high-speed data transmission. Photonics-assisted methods have the advantages of large processable bandwidth, low loss, and immunity to electromagnetic interference [6, 7], so many photonics-assisted SIC methods have been proposed in the past few years [8, 9].

To construct a reference signal, the adjustment of the delay and amplitude of the reference signal is required. The delay and amplitude adjustment in the photonics-assisted SIC methods can be implemented in the electrical domain or the optical domain. In the electrical domain, the delay and the amplitude can be adjusted by a tunable electrical delay line and a variable electrical attenuator [10, 11], respectively. The amplitude can also be adjusted by microwave voltage-controlled amplifiers [12]. However, a larger time delay achieved by the RF delay line will introduce more attenuation, and the cancellation bandwidth and depth will be limited by the frequency response flatness of the RF components. In the optical domain, the delay was commonly adjusted by the optical tunable delay line, and the amplitude was commonly adjusted by the variable optical attenuator [13-17]. Furthermore, the amplitude can also be adjusted by polarization controller [18], waveshaper [19], laser power [20], semiconductor optical amplifier (SOA) [21, 22], and modulator bias voltage [23-25], while the delay can also be tuned by bit-switched optical delay line [26], fiber Bragg grating (FGB) [27, 28], dispersion element [20], and SOA [21, 22]. However, all these delay and amplitude adjustment methods are performed in the analog domain, which requires additional devices. It is highly desirable that the amplitude and delay adjustment can be realized directly in the digital domain to simplify the system, which can also be better combined with the subsequent digital domain SIC.

In this paper, a photonics-based digital and analog RF SIC approach for IBFD systems is proposed, which can adaptively adjust the amplitude and delay of the reference signal in the digital domain. The amplitude setting of the reference signal is based on the power ratio of the system outputs when the received signal and the reference signal are input to the system separately. The delay value is searched by the cross-correlation and bisection method. The analog and digital SIC of the quadrature phase-shift keying (QPSK) modulated signals and the linearly frequency-modulated (LFM) signals are demonstrated. The analog cancellation depth is around 20 dB, and the total cancellation depth is more than 36 dB for the 2-Gbaud QPSK-modulated signals. For the LFM signals, the analog and total cancellation depths are around 19 dB and 34 dB, respectively.

## 2. Principle and experimental setup

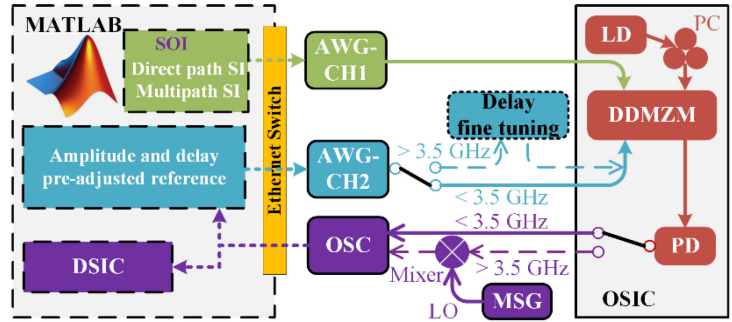

Fig. 1. Schematic diagram of the proposed RF SIC system. SOI, signal of interest; SI, self-interference; DSIC, digital self-interference cancellation; OSIC, optical self-interference cancellation; LD, laser diode; PC, polarization controller; DDMZM, dual-drive Mach-Zehnder modulator; PD, photodetector; AWG-CH, arbitrary waveform generator channel; MSG, microwave signal generator; LO: local oscillator signal; OSC, oscilloscope.

The schematic diagram of the proposed RF SIC system is shown in Fig. 1. A continuous-wave (CW) lightwave generated from a laser diode (LD, ID Photonics CoBriteDX1-1-C-H01-FA) with a wavelength of 1550.32 nm and a power of 15.5 dBm is injected into a dual-drive Mach-Zehnder modulator (DDMZM, Fujitsu FTM7937EZ200) via a polarization controller (PC). The DDMZM is biased at the quadrature transmission point. The upper RF port of the DDMZM is driven by the received signal, which includes the SI and the SOI, and is generated from the arbitrary waveform generator (AWG, Keysight M8195A, 64 GSa/s) channel 1. The lower RF port of the DDMZM is driven by the reference signal generated from the AWG channel 2 via an electrical delay line (Sage 6705). The optical signal from the DDMZM is detected in a photodetector (PD, Nortel networks PP-10G) with a responsivity of 0.88 A/W and a 3-dB bandwidth of 11 GHz. The analog SIC can be achieved when the reference signal matches the SI signal closely [12]. The electrical signal from the PD is directly sent to the real-time oscilloscope (OSC, R&S RTO2032, 3 GHz bandwidth, 10 GSa/s sampling rate) when the signal is with a relatively low frequency of less than 3.5 GHz. When the signal frequency is much higher than 3.5 GHz, it is firstly downconverted to the intermediate-frequency (IF) band using a mixer (M/A-COM M14A) and then sent to the OSC. The local oscillator signal injected into the mixer is generated from a microwave signal generator (MSG, Agilent 83630B), Due to the limited resolution of time adjustment through AWG, an electrical delay line in the dotted box in Fig. 1, which is optional and mainly used for the case with a very high signal frequency. The delay line can be removed if the signal frequency is low or more accurate time adjustment can be achieved in practical applications. The sampled waveforms are transmitted to a computer via an Ethernet switch and then processed in the digital domain using MATLAB, realizing two main functions: 1) the optimal values of the amplitude and delay of the reference signal are obtained and adjusted before the generation of the reference signal in the digital domain; 2) further digital domain SIC is realized through recursive least squared (RLS) or normalized least mean square (NLMS) algorithms to suppress the residual direct path SI and multipath SI. It should be noted that the AWG sampling rate and the OSC sampling rate are different. The received signal and the reference signal are generated by MATLAB with a sampling rate of 10 GSa/s. All the functions realized in the digital domain are implemented at a sampling rate of 10 GSa/s. After the processing in MATLAB, the received signal and the reference signal are resampled to 64 GSa/s and downloaded to the AWG via the Ethernet switch.

A proof-of-concept experiment according to the schematic diagram shown in Fig. 1 is conducted to verify the SIC ability of the proposed method. In the experiment, the received signal includes the SI and the SOI. The SI is composed of a direct path SI and a multipath SI. The delays and power attenuations of the multipath SI with respect to the direct path SI are 5 ns and 26 dB, respectively. The power of SOI is about 26 dB less than the power of direct path SI, and the bandwidth or baud rate of the SOI is 1/4 that of the SI. In the experiment, the resampled reference signal from the AWG is used for the analog domain SIC, whose amplitude and delay are properly adjusted in the digital domain directly. The amplitude setting of the reference signal is based on the power ratio of the system outputs when the received signal and the reference signal are input to the system separately. The delay of the reference signal is achieved in two steps, which include a cross-correlation step and a bisection method. The cross-correlation of the transmitted signal and the waveform from the OSC is obtained when the received signal and the reference signal without pre-adjusted delay are applied to the DDMZM. By using the cross-correlation, a rough estimate of the delay difference between the SI link and reference link is obtained, and the estimation accuracy of the delay difference is limited by the OSC sampling rate. For more precise delay adjustment, the delay of the resampled reference signal with 64 GSa/s sampling rate is further adjusted by using the bisection method according to the residual power after the analog SIC.

## 3. Experimental results and discussion

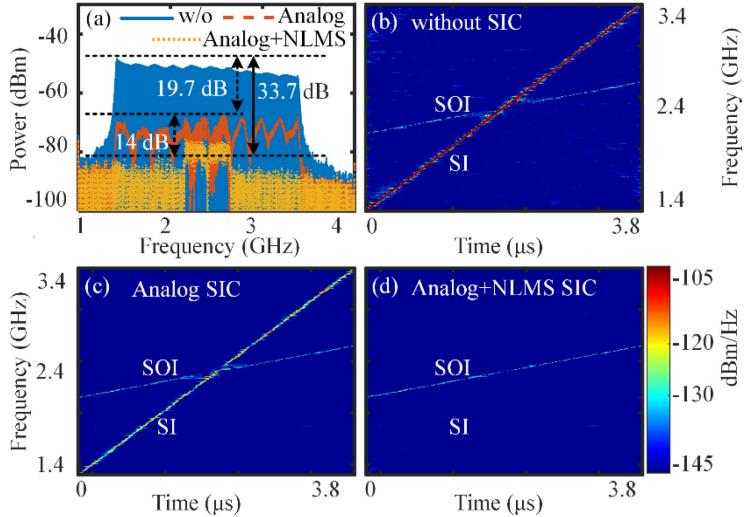

Fig. 2. (a) Electrical spectra of the LFM signal with and without SIC. The time-frequency diagram of the LFM signals (b) without SIC, (c) with only analog SIC, (d) with both analog and digital SIC.

Firstly, LFM signals for FMCW radars are used to verify the SIC performance of the system. To simplify the system and demonstrate the wideband operation, the signal is set to a relatively low frequency and sampled directly by the OSC without using the electrical mixer. Although in this case the waveform from the OSC is distorted due to the limited bandwidth of the OSC, the RF SIC is still implemented because we only manipulate the amplitude and delay of the references. In the experiment, the maximum amplitudes of the AWG outputs are both set to 1

V, which kept unchanged in the subsequent experiments. The cable lengths between the two arms of the DDMZM and AWG channel 1 and channel 2 are 2 m and 0.6m, respectively. To achieve the analog SIC, the reference signal amplitude is attenuated by multiplying a factor of 0.7868 in the digital domain, and the factor is acquired according to the power ratio of the received signal to the reference signal without attenuation. According to the previously mentioned two-step time delay determination method, the delay of the reference is set to 7.531 ns by adding a 482-point delay to the signal at a 64 Gs/s sampling rate.

Fig. 2(a) shows the electrical spectra of the LFM signals with and without SIC when the center frequency of the SI and the SOI are both 2.4 GHz and the bandwidths of the SI and the SOI are 2 and 0.5 GHz, respectively. As can be seen, the analog cancellation depth is around 19.7 dB. After digital SIC using the NLMS algorithm, an additional 14 dB cancellation is achieved and the total cancellation depth is more than 33.7 dB. It is noticed that the spectrum of the LFM signal is not flat, which is mainly caused by the limited bandwidth of the OSC. Furthermore, many ripples are observed when only analog SIC is enabled, which is mainly caused by the interference of the multipath SI and the residual direct path SI. Fig. 2(b) shows the time-frequency diagram of the LFM signal without SIC. In this case, the SI is very strong, while the SOI is with much lower power. Fig. 2(c) shows the time-frequency diagram of the LFM signal after analog SIC, where the SI is significantly suppressed compared with Fig. 2(b). Fig. 2(d) shows the time-frequency diagram of the LFM signal when both analog SIC and digital SIC are enabled, in which, the SI is further deeply suppressed and the SOI is dominated in the time-frequency diagram. It is also noticed from Fig. 2 that the power of the SOI does not have significant variations in the whole process.

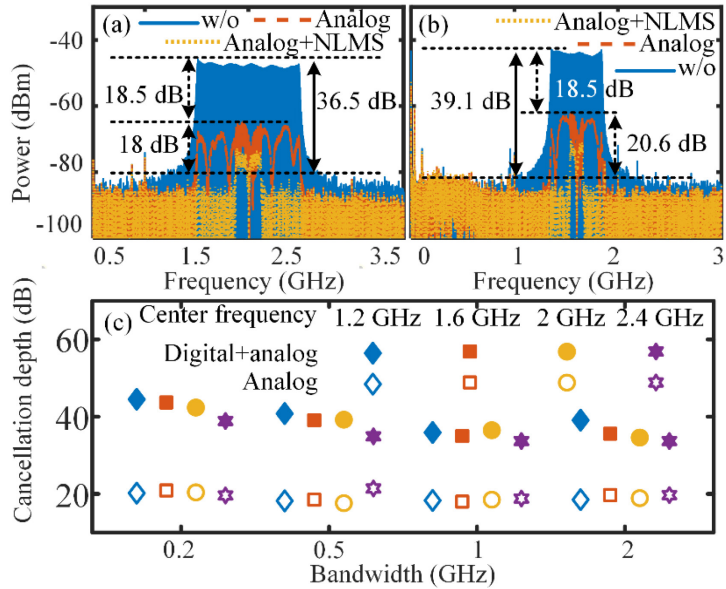

Fig. 3. Electrical spectra of the LFM signal with and without SIC when the center frequencies and bandwidths of the SI are (a) 2 GHz, 1 GHz, (b) 1.6 GHz, 0.5 GHz. (c) Cancellation depths of the LFM signal when the center frequencies are 1.2, 1.6, 2, and 2.4 GHz and the bandwidths are 0.2, 0.5, 1, and 2 GHz.

Then, LFM signals with different center frequencies and bandwidths are further used to verify the SIC performance of the system. The electrical spectra of the LFM signals with and without SIC are shown in Fig. 3(a) and (b). When the center frequency and bandwidth of the SI are 2 and 1 GHz, the analog cancellation depth is 18.5 dB. After digital SIC using the NLMS algorithm, an additional 18 dB cancellation is achieved. When the SI center frequency and bandwidth of the SI are decreased to 1.6 and 0.5 GHz, a total 39.1 dB cancellation depth is achieved with an 18.5 dB analog cancellation and a 20.6 dB digital cancellation. Fig. 3(c) shows the cancellation depths of the LFM signals when the SI center frequencies are 1.2, 1.6, 2, and 2.4 GHz, and the SI bandwidths are 0.2, 0.5, 1, and 2 GHz. The analog cancellation depths are all around 19 dB at different bandwidths and center frequencies. However, when the bandwidth or the center frequency of the SI increases, the total cancellation depth roughly decreases.

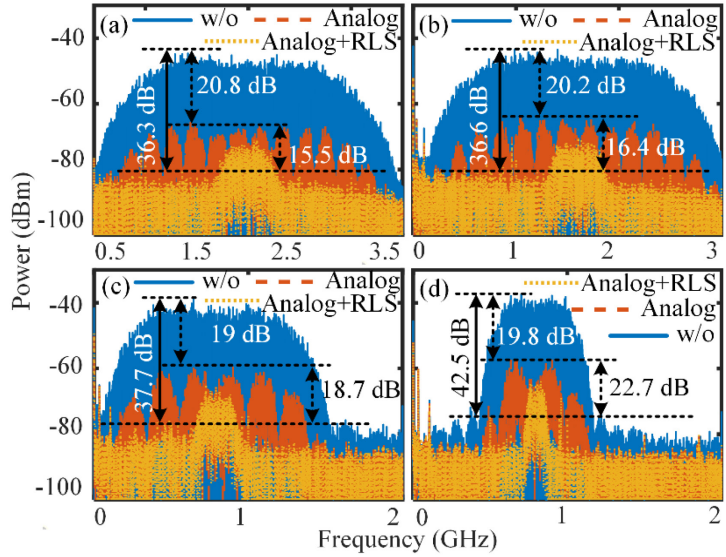

Fig. 4. Electrical spectra of the QPSK-modulated signals with and without SIC when the center frequencies and bandwidths of the SI are (a) 2 GHz, 2 Gbaud, (b) 1.6 GHz, 2 Gbaud, (c) 0.8 GHz, 1 Gbaud, (c) 0.8 GHz, 0.5 Gbaud.

Afterward, QPSK-modulated signals for IBFD systems are further used to verify the SIC performance of the system. The signal is also set to a relatively low frequency of less than 3.5 GHz. The electrical spectra of the 2-Gbaud QPSK-modulated signals with and without SIC are shown in Fig. 4(a) and (b). When the center frequency of the SI is 2 GHz, the analog cancellation depth is 20.8 dB. After digital SIC using the RLS algorithm, an additional 15.5 dB cancellation is achieved. When the center frequency of the SI is decreased to 1.6 GHz, the cancellation performance is very close to that of the 2 GHz center frequency. Fig. 4(c) and (d) show the electrical spectra of the QPSK-modulated signals when the center frequency of the SI is 0.8 GHz. The analog cancellation depth of the 1-Gbaud baud rate SI is 19 dB. After digital SIC using the RLS algorithm, an additional 18.7 dB cancellation is achieved. When the baud rate of the SI is decreased to 0.5 Gbaud, a total 42.5 dB cancellation depth is achieved with 19.8 dB analog cancellation and 22.7 dB digital cancellation. The SIC performance of the QPSK-modulated signals is similar to that of the LFM signals. Many ripples can also be observed when only analog SIC is enabled, which is also mainly caused by the interference of the multipath SI and the residual direct path SI.

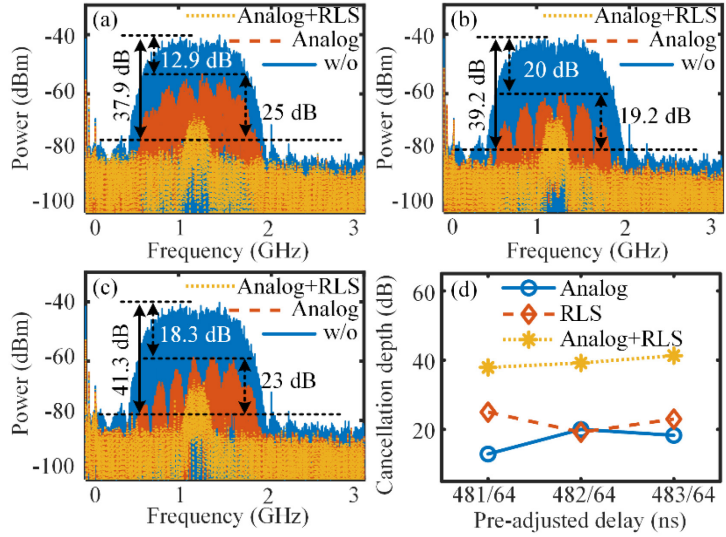

Fig. 5. Electrical spectra of the 1-Gbaud QPSK-modulated signal with and without SIC when the delay of the reference signal is (a) 7.516 ns, (b) 7.531 ns, (c) 7.547 ns. (d) Cancellation depth of the QPSK-modulated signal based on the three different delays.

The influence of different digital domain delays on the SIC performance is then studied, with the results shown in Fig. 5. The SI center frequency and baud rate are set to 1.2 GHz and 1 Gbaud, respectively. Fig. 5(a), (b), and (c) show the electrical spectra of the QPSK-modulated signal with and without SIC when the digital domain pre-adjusted delays are 7.516, 7.531, and 7.547 ns, respectively. The 7.516, 7.531, and 7.547 ns delays correspond to 481, 482, and 483 delay points with 64 GSa/s sampling rate. The cancellation depths based on the three different delays are shown in Fig. 5(d). The analog SIC performance is best at a 482-point delay, and the best delay point is acquired by the mentioned two-step delay determination method. When the delay is set to 483 points, the analog cancellation depth is a little worse than the delay is set to 482 points. When the delay is set to 481 points, the analog cancellation depth decreases significantly.

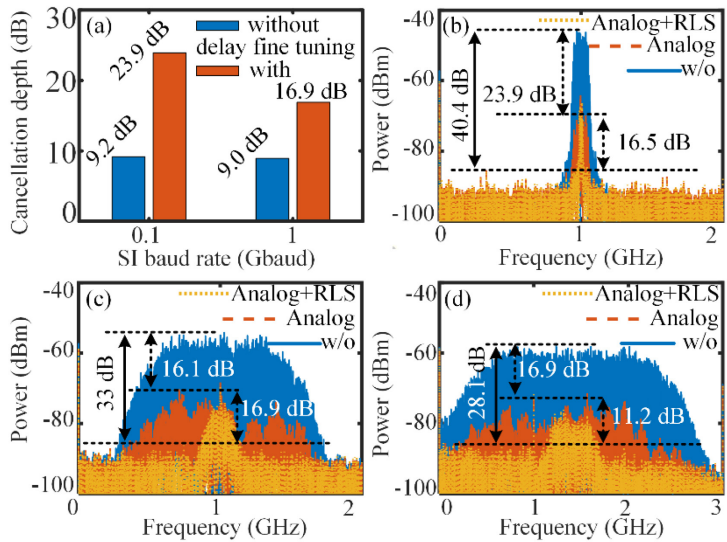

Fig. 6. (a) Analog cancellation depth with and without the delay fine tuning when the SI center frequency is 8 GHz. Electrical spectra of the down-converted IF signals with and without SIC when the SI symbol rate, the center frequencies of the IF signal and the LO signal are (b) 0.1 Gbaud, 1 GHz, 7 GHz, (c) 1 Gbaud, 1 GHz, 7 GHz, (d) 2 Gbaud, 1.4 GHz, 7 GHz.

Finally, the SIC performance of the system is further investigated when the signal is set to a relatively high frequency of more than 3.5 GHz. Owing to the OSC limited bandwidth, the mixer is used and the frequency of the LO is set to 7 GHz. The analog cancellation depth of SI with 7-GHz center frequency at different baud rates with and without the delay fine tuning are shown in Fig. 6(a). When the analog delay fine tuning is not applied, the analog cancellation depth is less than 10 dB, which is limited by the resolution of time adjustment through AWG. When the delay fine-tuning using the electrical delay line is implemented, the analog cancellation depths are greatly improved. Fig. 6(b), (c), and (d) show the electrical spectra of the QPSK-modulated signals with and without SIC when the analog delay fine tuning is applied by the electrical delay line. When the SI center frequency is 8 GHz and the baud rate is 0.1 Gbaud, the analog cancellation depth is 23.9 dB. After digital SIC using the RLS algorithm, an additional 16.5 dB cancellation is achieved. When the baud rate of the SI is increased to 1 Gbaud, the analog cancellation is decreased to 16.1 dB and there is no significant change in the digital domain SIC. When the baud rate of the SI is increased to 2 Gbaud and the center frequency of the SI is changed to 8.4 GHz. The digital cancellation is decreased to 11.2 dB and there is no significant change in the analog SIC.

## 4. Conclusion

In summary, we have demonstrated a photonics-based digital and analog RF SIC approach for IBFD communication systems and FMCW radar systems. The key significance of the work is that the amplitude and delay of the reference signal in the photonics-based RF SIC system are matched in the digital domain instead of the analog domain, which can reduce the cost and simplify the RF SIC structure, in addition to providing a better combination with the subsequent digital domain SIC. An experiment is carried out to verify the cancellation performance of the system. The analog cancellation depth is around 20 dB, and the total cancellation depth is more than 36 dB for the 2-Gbaud QPSK-modulated signals. For the LFM signals, the analog and total cancellation depths are around 19 dB and 34 dB, respectively. The proposed approach provides a new perspective of the delay and amplitude adjustment for the photonics-assisted SIC systems.

**Acknowledgements**
This work was supported by the National Natural Science Foundation of China [grant number 61971193]; the Natural Science Foundation of Shanghai [grant number 20ZR1416100]; the Open Fund of State Key Laboratory of Advanced Optical Communication Systems and Networks, Peking University, China [grant number 2020GZKF005]; and the Science and Technology Commission of Shanghai Municipality [grant number 18DZ2270800].